\documentclass{aa}
\usepackage{graphicx}
\begin{document}
\title{Detection of a redshift 3.04 filament
             \thanks{
Based on observations collected at the European Southern
Observatory, Paranal, Chile (ESO Programme 64.O-0187)} }

\author{P. M\o ller
        \inst{1}
        \and
        J.U. Fynbo
        \inst{1}
        }

\offprints{P. M\o ller}
\institute{
         European Southern Observatory, Karl-Schwarzschild-Stra\ss e 2,
         D-85748, Garching by M\"unchen, Germany\\
         \email{pmoller@eso.org \& jfynbo@eso.org}
         }

\date{Received ; accepted }

\abstract{The filamentary structure of the early universe has until now
only been seen in numerical simulations. Despite this lack of direct
observational evidence, the prediction of early filamentary structure
formation in a Cold Dark Matter dominated universe has become a paradigm
for our understanding of galaxy assembly at high redshifts. Clearly
observational confirmation is required. Lyman Break galaxies are too
rare to be used as tracers of filaments and we argue that to map out
filaments in the high z universe, one will need to identify classes of
objects fainter than those currently accessible via the Lyman Break
technique. Objects selected via their Ly$\alpha$ emission, and/or
as DLA absorbers, populate the faintest accessible part of the high
redshift galaxy luminosity function, and as such make up good candidates
for objects which will map out high redshift filaments. Here we
present the first direct detection of a filament (at z=3.04)
mapped by those classes of objects. The observations are the deepest
yet to have been done in Ly$\alpha$ imaging at high redshift, and
they reveal a single string of proto-galaxies spanning about 5 Mpc
(20 Mpc comoving). Expanding the
cosmological test proposed by Alcock \& Paczy{\'n}ski (1979), we outline
how observations of this type can be used to determine
$\Omega_{\Lambda}$ at z=3.
}

   \maketitle
%

\section{Introduction}
For the past three decades, following the first prediction of
pancakes by Zel'dovich 1970, simulations have been ahead of the
observations when it comes to describing the first structures to form
at high redshift.
Numerical simulations of structure formation based on the Cold Dark
Matter scenario predict that the first large scale structures to form
are voids and filaments (Klypin \& Shandarin 1983;
White et al. 1987; Evrard et al. 1994; Rauch et al. 1997).
In this picture any sightline through
the early universe will intersect a large number of voids
and filaments, and the variation in the density of neutral hydrogen
along the sightline is then observed as the Lyman forest
(Petitjean et al. 1995; Theuns et al. 1998).
Growth of density fluctuations along the filaments will lead to
formation of lumps of cold, self-shielding gas and those are
identified, in the simulations, as regions of star formation
(Katz et al. 1996; Haehnelt et al. 2000).
Because of the high column density of neutral hydrogen in such clouds,
they are identified observationally as strong absorbers and the
strongest as Damped Ly$\alpha$ Absorbers (DLAs). By poking random
sightlines through a virtual universe one may simulate observations,
and a given model universe will hence predict a specific
correlation between DLA systems and the galaxies hosting the DLAs
(Katz et al. 1996). Comparison to real observations of DLA
galaxies (M{\o}ller \& Warren 1998) has shown that there
is very good agreement between
observations and simulations. This agreement is encouraging, but it
would be of great interest if one could observationally map out the
actual high redshift filaments. Apart from being a fundamental test of
the validity of the simulations and the structure formation
paradigm they represent, knowing the scale size of the
filaments at different redshifts will help constrain both cosmological
parameters and the mechanisms of galaxy formation.

Unfortunately such a map cannot be constructed via absorption studies,
because there is no sufficiently tight mesh of background quasars
available. The only way to proceed is to attempt to find enough
centers of star formation to map out the filament in its own light, as
has been done at low redshift (z$<$0.2) using galaxy surveys
(Gregory \& Thompson 1978; de Lapparent et al. 1991; Bharadwaj et al.
2000). At low redshifts filaments remain, but many of the galaxies have
drained into the filament nodes where todays galaxy clusters are
formed. In order to identify objects in a high redshift filament one
might first try a search for Lyman Break Galaxies (LBGs,
Steidel \& Hamilton, 1992).
Unfortunately only relatively bright galaxies can be found with this
technique, and sparse sampling of the filamentary structure does not
allow it to be seen easily. However, both DLA galaxies and galaxies
selected for their Ly$\alpha$ emission, sample the high redshift
galaxy population at much fainter magnitudes than do the LBGs, and
should therefore give a better sampling of the high redshift
structure
(Fynbo et al. 1999; Haehnelt et al. 2000; Fynbo et al. 2001 (FMT2001)).
This has recently been independently confirmed,
by deep narrow band Ly$\alpha$ imaging in a field of known LBGs, which
reveals about a factor of 6 more candidate Ly$\alpha$ galaxies than
LBGs (Steidel et al. 2000).
A hint that filamentary structure is indeed present, comes
from a few cases where narrow-band Ly$\alpha$ imaging of
DLAs was attempted, and one or two additional neighbour
Ly$\alpha$ emitters were found. In particular it has been demonstrated
that when three DLA and/or Ly$\alpha$ sources are detected in
a field, they show a marked tendency to be aligned
(M{\o}ller \& Warren 1998).

\begin{table}
\begin{center}
\caption{Redshifts and positions of seven Ly$\alpha$ emitters and a
Ly$\alpha$ absorber in the field of Q1205--30. The positions are
given relative to the quasar coordinates:
12:08:12.7, -30:31:06.10 (J2000.0). The uncertainty on the redshifts
is 0.0012 (1$\sigma$).}
\begin{tabular}{@{}lrrl}
\hline
Object & $\Delta$RA (arcsec) & $\Delta$decl. (arcsec) & redshift \\
\hline
S7  & -143.3$\pm$0.6 &  41.9$\pm$0.2 & 3.0402  \\
S8  & -141.5$\pm$0.6 &  59.7$\pm$0.2 & 3.0398  \\
S9  & -124.6$\pm$0.5 &  63.4$\pm$0.2 & 3.0350  \\
S10 & -119.9$\pm$0.5 &  59.8$\pm$0.2 & 3.0353  \\
S11 &  -77.8$\pm$0.3 &   0.9$\pm$0.1 & 3.0312  \\
S12 &  -43.9$\pm$0.2 &  54.4$\pm$0.2 & 3.0333  \\
S13 &   68.3$\pm$0.3 & -52.1$\pm$0.2 & 3.0228  \\
abs &    0.0         &   0.0         & 3.0322  \\
\hline
\label{redtab}
\end{tabular}
\vspace{-0.8cm}
\end{center}
\end{table}

\section{Objects in the field of \object{Q1205-30}}
In Fynbo et al. 2000 we reported on a new very deep,
$1.1 \times 10^{-17}$ erg s$^{-1}$ cm $^{-2}$ ($5\sigma$), narrow band
search for Ly$\alpha$ emitters in a 28 arcmin$^2$ field centred on the
quasar Q1205--30. The filter was tuned to the wavelength of a
strong Ly$\alpha$ absorption line in the quasar spectrum.
On March 4--5, 2000 we used the ESO Very Large Telescope on Paranal to
obtain spectra of the candidate Ly$\alpha$ emitters in the field.
Details of the spectroscopic observations and data reductions have been
submitted for publication (FMT2001), but the results are summarised in
Table 1 where we list the position and redshift of each of the
seven confirmed Ly$\alpha$ emitting objects. In the table we also
give the coordinates and redshift of the absorber.

\begin{figure}
  \centering
  \includegraphics[width=8cm]{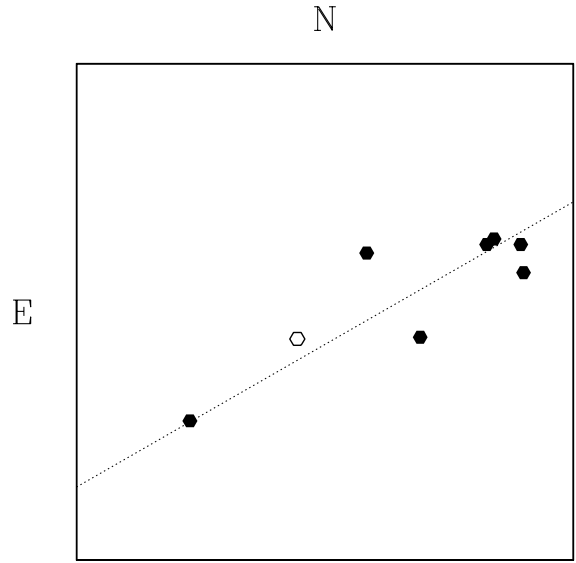}
  {\vspace{-0.6cm}}
    \caption{
    Position of the eight objects on the CCD. In this and all
    following figures solid hexagons mark the Ly$\alpha$ emitters, the
    open hexagon marks the absorber.}
  \label{field}
\end{figure}

In Fig.~1 we show a sky-coordinate map of the field. While there may
here be a hint of an elongated structure running from the SE
corner of the CCD frame towards the NW, the statistical evidence for
a non-random distribution is weak.
However, this may be because we are looking almost along a filament.
If this was the case, we would expect the redshifts to change
monotonically along the structure suggested by the dotted line in
Fig.~1. This should give rise to a strong correlation between redshift
and position on the CCD.
We performed two tests to assess this correlation, one Monte Carlo
and one analytical:

\begin{figure}
  \centering
  \includegraphics[width=9cm]{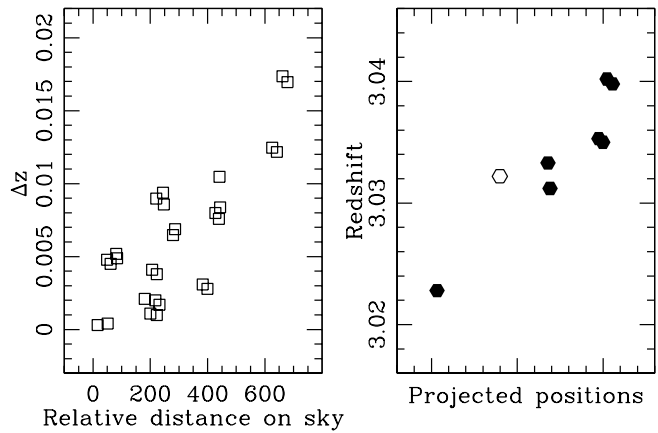}
    \caption{
    {\bf Left:}
    Redshift difference plotted against relative distance on the sky (in
    CCD pixels). The correlation coefficient is $r_s = 0.73$. In random
    distributions the correlation parameter is worse than that in
    99.7\% of all cases.
    {\bf Right:}
    Redshift plotted against projected 1D distance (positions projected
    onto the dotted line in Fig.~1).}
  \label{correl}
\end{figure}

\begin{figure*}
  \centering
  \includegraphics[width=16cm]{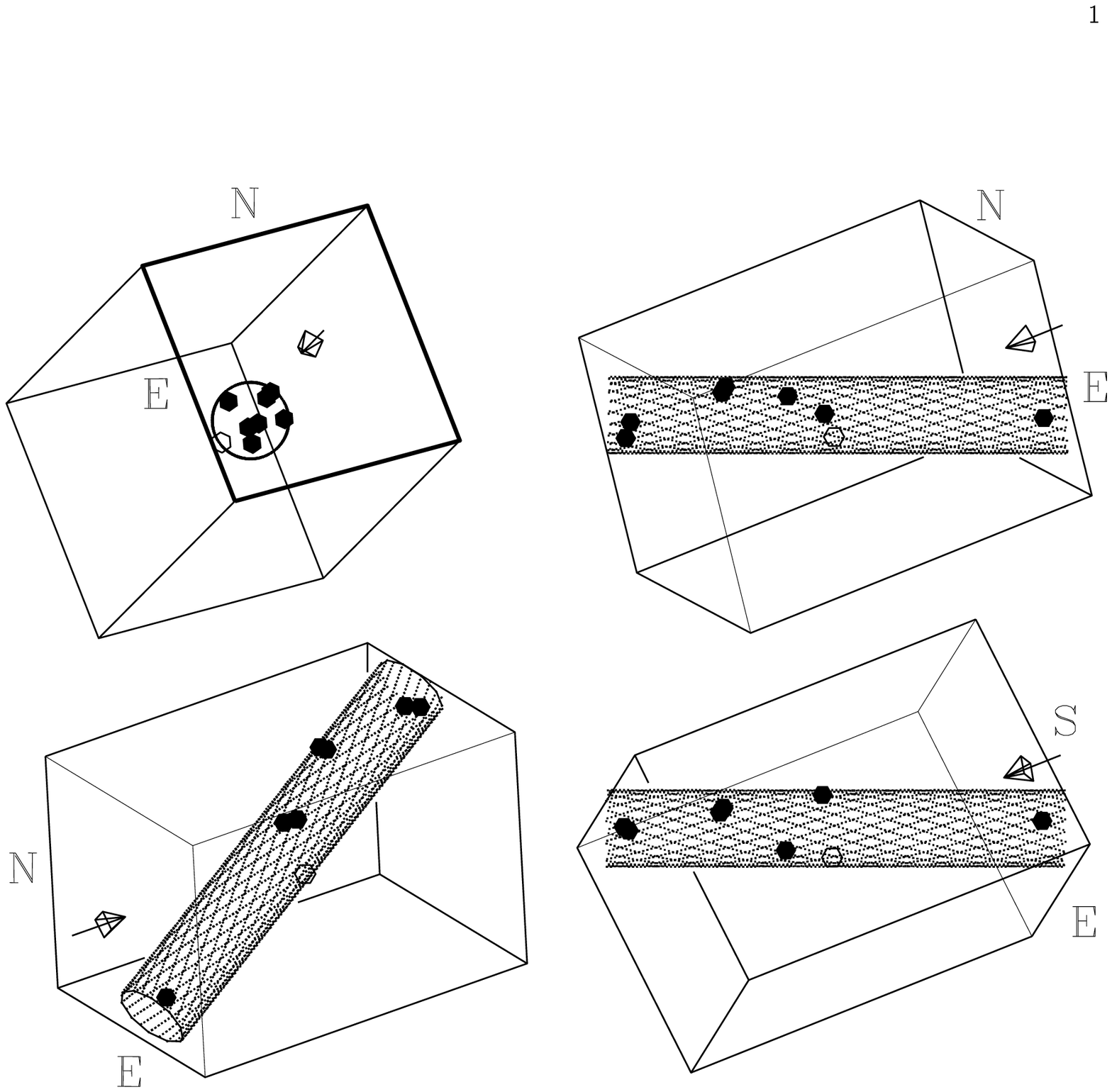}
  {\vspace{-4.5cm}}
    \caption{
    3D distribution of the eight objects seen from 4 different
    viewing-angles.
    In each of the figures the 3D arrow points in our viewing
    direction into the sky, and the spiral pattern maps out a cylinder
    with diameter 400 kpc (see Table 2). The box marks the volume of
    space observed with our narrow-band Ly$\alpha$ filter.
    {\bf Top left:}
    Here we have rotated the view to look along the filament. The thick
    lines mark the front "entrance window" of the box (corresponding to
    our CCD image).
    {\bf Top right:}
    The box is here rotated 90 degrees to the right, hence viewing the
    filament from the left side compared to the end-on view.
    {\bf Bottom right:}
    Same as top right but rotated 90 degrees around the filament to give
    a view of the filament as seen from ``above'' the view in top left.
    {\bf Bottom left:}
    View from a random angle to give an impression of the 3D structure.
    }
  \label{3D}
\end{figure*}

1) For each of the eight objects listed in Table 1 we measured the
projected distance on the sky, as well as the difference in redshift,
to each of the other seven objects. Those measurements are reproduced
in Fig.~2(left). A Spearman correlation test of
the data points gives a correlation coefficient of $r_s = 0.73$.
We then produced $3\times 10^6$ realizations of a random distribution
of eight objects in the observed volume, keeping the position and
redshift of the absorber fixed. In only 0.3\% of the simulations did
the same test give a similar or better correlation. A more fair
comparison is probably to compare against a random distribution with
an imposed two-point correlation function as reported by
Giavalisco \& Dickinson 2001
($\xi(r)=(r/r_0)^{-\gamma}$, $r_0=1.1h^{-1}{\rm Mpc}$, $\gamma = 2.2$).
In this case 0.4\% of
the simulations gave a similar or better correlation.

2) Encouraged by this tentative suggestion of a filament,
we proceeded to project the position of each object on the CCD onto
the line drawn in Fig.~1. In Fig.~2(right) we plot the redshift of each
object against the projected position. A Spearman correlation test
now gives $r_s = 0.881$ for eight objects. In this case we are dealing
with statistically independent data points, which confirms the 
correlation with a probability of 99.82\%.

These tests confirm that the eight objects form a tight
string-like structure in redshift space.
Assuming that the redshifts are caused by Hubble flow, this structure
takes the form of a 3D filament similar to those in numerical
simulations. The objects are fainter and much
more common than the typical LBGs and they presumably represent an
earlier evolutionary state than the brighter LBG galaxies. Evidence for
this is found in the fact that only 20\% of the LBG population has
Ly$\alpha$ in emission, and as a class they show evidence for internal
dust absorption suggesting high metallicity (Steidel et al. 2000).
In contrast, the numerous faint objects with Ly$\alpha$ in
emission have much lower star formation rates than the LBGs and are
essentially dust free (FMT2001). In the simulations the
final fate of several such objects is to merge into one single galaxy,
and for this reason we shall here refer to them as
``Ly$\alpha$ Emitting Galaxy-building Objects'' (LEGOs).

\section{The Q1205-30 filament at z=3.04}
In Fig.~3 we show the 3D structure of the volume of space we are
sampling from 4 different viewing-angles. Our sample inside this
volume is complete down to a limiting Ly$\alpha$ flux of
$1.1 \times 10^{-17}$ erg s$^{-1}$ cm $^{-2}$ ($5\sigma$). The redshift
distribution of the LEGOs is consistent with a random distribution
when one allows for the selection defined by the filter transmission
curve (FMT2001), and we hence conclude that the
distribution along the filament is consistent with a random
distribution. In other words, we have not detected the ends of the
filament, and it may extend further in either direction.

In each of the four figures in Fig.~3 we have enclosed the LEGOs in a
hollow ``tube'' which is meant to illustrate the surface of the
filament. The same filament-surface is shown in all the figures.
In particular in Fig.~3(top left) where the filament is
viewed from the end, the tube takes the form of a circle.
To define the scales of the filament we have
chosen three parameters: The length, the radius of the tube and the
rms of the projected distribution of the objects around the center
of the filament when the filament is viewed from the end. The
actual values of those parameters will depend on the
chosen cosmology. In Table~2 we list the results for a representative
selection of cosmologies. Note that to represent the length we have
chosen the largest distance between two detected LEGOs. As
discussed above this is in reality a lower limit to the length as we
have not yet determined where the endpoints are. The comoving length
is hence found to be minimum 20 Mpc, in good agreement with recent
predictions (20-25$h^{-1}$, Demia\'nski \& Doroshkevich 1999).

\begin{table}
\begin{center}
\caption{Filament parameters for different cosmologies. In all
cases have we used H$_{\rm o}$ = 65 km s$^{-1}$ Mpc$^{-1}$. We
defined the length as the distance between the two outhermost
LEGOs.}
\begin{tabular}{@{}cccc}
\hline
Model & rms (kpc) & Radius (kpc) & Length (kpc) \\
\hline
$\Omega = 0.2$, $\Omega_{\Lambda} = 0.0$ &  280 & 400 & 4400 \\
$\Omega = 1.0$, $\Omega_{\Lambda} = 0.0$ &  180 & 260 & 2800 \\
$\Omega = 0.3$, $\Omega_{\Lambda} = 0.7$ &  280 & 400 & 4800 \\
$\Omega = 0.1$, $\Omega_{\Lambda} = 0.9$ &  390 & 560 & 7700 \\
\hline
\label{filtab}
\end{tabular}
\vspace{-1cm}
\end{center}
\end{table}

In an $\Omega = 0.2$, $\Omega_{\Lambda} = 0.0$ cosmology the volume
sampled by our observations is 30 Mpc$^3$. The filament
fills only about 7\% of this volume, consistent with a picture of
the early universe where the non-linear processes leading to the
formation of proto-galaxies are confined within thin filaments.

\section{High z filaments as cosmological probes}
It is interesting to note that five of the seven LEGOs remain
undetected in the individual broad-band images (FMT2001), which reach
limiting continuum AB magnitudes of I=26.9 and B=27.7 (2$\sigma$ upper
limits). Our argument at the outset that searches
for LBGs will not detect this type of filament is
clearly vindicated. Moreover most of the LEGOs reported here
have Ly$\alpha$ fluxes very close to our
detection limit. Hence, a slightly shallower Ly$\alpha$ search
would likewise have failed to detect the filament. Finally, had the
orientation of the filament not been so closely aligned with our line
of sight, then only a fraction of the LEGOs would have fallen within
the CCD field of view, again making a detection impossible.
In this particular case only the fortunate near-alignment and the
close-to-critical detection limit made the detection possible.
Ly$\alpha$ searches to the same or slightly deeper detection limits but
over larger fields should easily pick out several such filaments,
thereby making it possible to map the filamentary structure of a
significant volume of the early universe.

This opens up an interesting perspective as it will provide an
independent way to determine the cosmological constant
$\Omega_{\Lambda}$. The conversion of the observed volume into
proper coordinates depends on the chosen cosmological model. 
The effect of using a non-zero $\Omega_{\Lambda}$ is to significantly
stretch the volume along the sightline, thereby causing the ``length
to radius ratio'' of filaments to be different for filaments seen
from the side and filaments seen end-on. This is a new realisation
of the test proposed by Alcock \& Paczy{\'n}ski 1979, who considered
the elongation of an idealized spherically symmetric distribution of
emitters. In the case of emitters distributed in filaments the test
becomes somewhat more complex because of the lack of spherical symmetry,
and one will need to consider the statistics of a sample of filaments.
At the same time, however, the lack of spherical symmetry offers
several independent tests. We consider three tests to be of practical
use. First one may consider the distribution of inclination angles
of filaments which should be isotropic,
secondly the ratio of the rms to the length (as given in Table 2) which
must be independent of filament inclination angle, and third the
cross-section of the filaments which must in the mean be circular
for all inclination angles. In practice one should determine the value
of $\Omega_{\Lambda}$ which provides the best fit to all three
tests simultaneously. As already pointed out by Alcock \&
Paczy{\'n}ski, random motion of objects with respect to the
Hubble-flow will add some noise to any such test. We shall return to
a detailed discussion of this in a forthcoming paper.

This new set of $\Omega_{\Lambda}$ tests is useful in the redshift range
$z=2$--4, is within the capability of current instrumentation and is
independent of the successful SN
(Riess et al. 1998; Perlmutter et al. 1999) and weak-lensing
(Maoli et al. 2001)
cosmological tests which currently can be used out to $z \approx 1$.
With the complement of large telescopes currently available, we
predict rapid progress in this field.

\begin{acknowledgements}
We are grateful to B. Thomsen for permission to use results prior to
their publication and to S.D.M. White for stimulating discussions and
many useful comments on an early draft of this manuscript.
\end{acknowledgements}

\end{document}